\documentclass[superscriptaddress,twocolumn,showpacs,prb,floatfix]{revtex4}
\usepackage{graphicx}
\usepackage{tabularx}
\begin{document}

\title{The Bimodal Ising Spin Glass in dimension three : Corrections to scaling}
\author{K.~Hukushima}
\affiliation{
Department of Basic Science, University of Tokyo, Tokyo, 153-8902, Japan}
\author{I.~A.~Campbell}
\affiliation{Laboratoire des Collo\"ides, Verres et Nanomat\'eriaux,
Universit\'e Montpellier II, 34095 Montpellier, France}

\begin{abstract}
Equilibrium numerical data on the three dimensional bimodal ($\pm J$)
 interaction Ising spin glass up to size $L=48$ show that corrections to
 scaling, which are known to be strong, behave in a non-monotonic manner
 with size. Extrapolation to the infinite size thermodynamic limit is
 difficult; however the large $L$ data indicate that the ordering
 temperature $T_c$ lies significantly higher than the values which have
 been estimated from previous numerical work limited to smaller
 sizes. 
In view of the present results it is at the least premature to  
conclude that the three dimensional bimodal and Gaussian Ising spin glasses lie in  
the same universality class.
\end{abstract}

\pacs{75.50.Lk, 75.40.Mg, 05.50.+q}
\maketitle

\section{Introduction}

Renormalization group theory (RGT) provides an explanation of the physical origin of the critical exponents and of the universality classes for standard second order transitions which is one of the most remarkable achievements of statistical physics.
The universality rules state that the critical exponents depend only on
a small number of basic parameters~\cite{ma:76}, essentially the 
dimension of space $D$, the range of interaction,  and the number of
order parameter components $n$. Physically, the critical parameters should not depend on the
details of the short range interactions. The few known and well
understood exceptions to universality concern mainly rare marginal
cases, such as certain regularly frustrated spin systems in two
dimensions where critical exponents vary continuously with the value of
a control parameter. It should also be kept in mind that if the specific
heat exponent $\alpha$ is positive for pure ferromagnets, 
disorder induces another universality class\cite{HarrisLubensky:74}.  

The glass transition in structural glasses has long raised fundamental
questions. It has been hoped that the study of spin glass (SG)
transitions would give insight into the basic physics of glassy
ordering. In particular,  a natural assumption to make is that
universality rules hold in SGs, with exponents which are different from
those of ferromagnets but which do not depend on the detailed form of
the interactions between spins. 

For technical and conceptual reasons the SG transition should be much easier to study than structural glass freezing but it has turned out that SG transitions also pose major problems. It has proved impossible so far to extend RGT satisfactorily to dimensions below the upper critical dimension, so information on transitions comes from experiment and
the large scale numerical simulations which can be carried out on
SGs. Numerous numerical studies on Ising Spin Glasses (ISGs) have
concluded that universality holds also for these systems. However,
determining transition temperatures and critical exponents to high
precision at SG transitions through simulations remains a notoriously
difficult task. Problems that are intrinsic to the numerical simulations
include agonizingly slow equilibration rates near criticality and the
need to average over a large number of microscopically inequivalent
samples. It is helpful to analyze the numerical data using appropriate
scaling variables and scaling rules~\cite{campbell:06}, and above all
it is essential to allow adequately for corrections to finite size
scaling (FSS). 

Here we address once again the case of the Ising spin glass with
bimodal ($\pm J$) interactions in dimension three (3D), a system which has
already been the subject of many numerical studies. Recent careful
measurements~\cite{kawashima:96,ballesteros:00,katzgraber:06,hasenbusch:08},
where samples of sizes up to $L=24$, $20$, $24$ and $L=28$ respectively were
studied, all give estimates of the critical temperature consistent with
$T_c \sim 1.12(1)$.  In the present work equilibrium measurements have
been made for sample sizes up to twice as large as those studied in most
of the previous work. The data demonstrate that in this system the
influence of finite-size corrections to scaling extends to very large
sizes, strongly affecting the thermodynamic limit estimate of the
ordering temperature and therefore those of the critical exponents in
this system. An analysis of these equilibrium data allowing for the
corrections suggests a higher critical temperature, compatible with
estimations from a ``dynamic''
approach~\cite{bernardi:96,mari:01,pleimling:05}. As a corollary the
critical exponents will be significantly different from those obtained
with $T_c\sim 1.12$, as in Ref.~\onlinecite{kawashima:96,ballesteros:00,katzgraber:06,hasenbusch:08}. 

Our principal conclusion is that the present large size data for the
3D Bimodal ISG taken together with published data for the 3D Gaussian
ISG where corrections appear to be much
weaker (Ref.~\onlinecite{katzgraber:06}) cannot be read as evidence that
the two systems lie in the same universality class.

\section{Measurements and analysis}

We measure as usual 
the reduced spin-glass susceptibility $\chi(T,L)$ 
defined as 
\begin{equation}
 \chi(T,L) = \frac{1}{L^3}\left\langle\left(\sum_i S_i^{(1)}S_i^{(2)}\right)^2
			  \right\rangle , 
\end{equation}
where $\langle\cdots\rangle$ represents thermal and disorder average,
and the upper suffix denotes two real replicas of the model with
identical bond configurations. Using the spin overlap given by 
\begin{equation}
 q = \frac{1}{L^3}\sum_i S_i^{(1)}S_i^{(2)}, 
\end{equation} 
the spin-glass susceptibility is also expressed as $\chi=L^3\langle
q^2\rangle$. 
We also measure finite-size correlation length~\cite{cooper:82}, 
\begin{equation}
 \xi(T,L) =
  \frac{1}{2\sin(\pi/L)}\left[\frac{\tilde{\chi}(\mathbf{0})}{\tilde{\chi}(\mathbf{k}_{\rm
			 min})}-1\right]^{1/2}, 
\end{equation}
where $\tilde{\chi}(\mathbf{k})$ is the Fourier transformed spin-glass
 susceptibility with wave vector $\mathbf{k}$ and 
 $\mathbf{k}_{\rm min}=(2\pi/L,0,0)$. 
This becomes identical to the second moment
correlation length in the thermodynamic limit $L \gg \xi(\infty,T)$. 

Published estimates of critical exponents on specific ISGs have varied
widely (see the list in Ref.~\onlinecite{katzgraber:06}) and it is of
interest to examine the reasons for this dispersion. In most work FSS
analyses are carried out on measurements made at equilibrium on small to
moderate sized samples; the sizes $L$, the number of independent
samples, and the equilibration times were necessarily restricted by
computer time limitations. The ordering temperature $T_c$ is generally
estimated by the intersection temperature of curves for the Binder
parameter given by 
\begin{equation}
g(L,T)=\frac{1}{2}\left(3-\frac{\langle q^4\rangle}{\langle q^2\rangle^2}\right)
\end{equation}or of the normalized correlation
length $\xi(L,T)/L$  as these two observables become independent of size
at the ordering temperature, to within FSS corrections. This caveat is
vital because for the sizes which can be equilibrated in practice FSS
corrections are often not negligible and it is hard to extrapolate
precisely from data on a set of finite $L$ samples to the thermodynamic
(infinite $L$) limit because, in contrast to the situation in standard
ferromagnets, there are no firm guidelines as to the form of the
corrections at and close to criticality. We will return to this crucial
point below. 

Once $T_c$ has been estimated the critical exponents $\nu$ and $\eta$ (with $\gamma=(2-\eta)\nu$) have generally been estimated through a standard FSS analysis on the Binder parameter, the normalized correlation length, and the reduced susceptibility $\chi(L,T)$ :
$g(L,T) = F_{g}[L^{1/\nu}(T-T_c)]$, $\xi(L,T)/L =
F_{\xi}[L^{1/\nu}(T-T_c)]$ and
$\chi(L,T)=L^{2-\eta}F_{\chi}[L^{1/\nu}(T-T_c)]$. These rules, which
have been taken over directly from the usage in ferromagnet studies,
implicitly assume that the appropriate temperature scaling variable is
$[(T-T_c)/T_c]$. In a ferromagnet the coupling strength is $J$ and so
reduced field and temperature scales are $H/J$ and $T/J$. In a SG the
effective coupling strength parameter is $\langle J^2\rangle$ not $J$
and it is the non-linear susceptibility, not the linear susceptibility,
which diverges at criticality. Hence the appropriate ``field'' and
``temperature'' variables are $H^2/\langle J^2\rangle$ and $T^2/\langle
J^2\rangle$ rather than $H/J$ and $T/J$ (see
Refs.~\onlinecite{fisch:77,singh:86,klein:91,daboul:04}). We have shown
\cite{campbell:06} that it is useful for SGs to employ ``extended
scaling'' rules which (writing $\beta = 1/T$) make the leading
approximations  
\begin{equation}
\xi(\infty,\beta) \sim \beta\left[1-(\beta/\beta_c)^2\right]^{-\nu}
\end{equation}
and
\begin{equation}
\chi(\infty,\beta) \sim [1 - (\beta/\beta_c)^2]^{-\gamma}
\end{equation}
These reduce to the standard forms in the limit $(T-T_c)/T_c \ll 1$, and
they also lead to the functionally correct forms in the high temperature
limit. They become exact at all $T>T_c$ in the high dimension limit,
but correction terms are to be expected in finite dimensions. When these
scaling expressions are used, estimates for the exponents derived from
scaling data on the different observables over wide temperature ranges
are considerably more consistent than when the traditional rules are
used to analyze the same data sets (see Ref. \onlinecite{katzgraber:06} where
the two analyses are compared). The traditional expressions reduce to
the SG expressions very close to criticality, but if they are used to
scale data covering a wider range of $T$ they tend to strongly bias
estimates of $\nu$. Consequently, previous works did not provide a
consistent estimation of $\nu$ that should be independent of physical
quantities. This remark in itself explains why estimates of
$\nu$ in older publications were systematically very low compared to
more recent values\cite{katzgraber:06}. 

There remains the basic technical problem of determining of an adequate
sample size such that corrections to FSS have become negligible. Sizes
$L$ cannot be increased at will because equilibration at and near $T_c$
becomes rapidly more difficult with increasing $L$, leading to
escalating cost in computer time. We have chosen a strategy which
sidesteps this problem by appealing to the finite-size ratio scaling
approach introduced by Caracciolo {\it et al} ~\cite{caracciolo:95}, and
already applied to the 3D bimodal ISG by Palassini and Caracciolo\cite{palassini:99}. For each observable $Q(L,T)$ the ratio
$Q(sL,T)/Q(L,T)$ is plotted against the normalized correlation length
$\xi(L,T)/L$. In practice we scale with $s=2$ and so plot the ratios
$Q(2L,T)/Q(L,T)$, with $Q=\xi$, $\chi$ and $g$. In the limit of
sufficiently large $L$ where finite size corrections to scaling have
become negligible, for each observable $Q(L,T)$ all the scaled data must
fall on a single curve characteristic of the universality class of the
system\cite{caracciolo:95}. In this limit $\xi(2L,T_c)/\xi(L,T_c)$ and
$g(2L,T_c)/g(L,T_c)$ are equal to 2 and 1 respectively when $\xi(L,T)/L$
is equal to its critical value. Instead of only using the intersections
of $\xi(L,T)/L$ and $g(L,T)$ curves at and near $T_c$ to judge whether
the FSS-correction-free limit has been reached, we can use the entire
family of scaling curves. The important point is that if already the
scaled data do not lie on a single $L$-independent curve at temperatures
somewhat above $T_c$, then finite size corrections to scaling are still
important for the sizes studied and the thermodynamic limit has not yet
been reached. Because of the high value of the dynamical exponent
$z(T_c)$ in ISGs ($z(T_c)$ is typically $\sim 6$ in 3D) it is
technically much faster to equilibrate large samples down to say
$1.2T_c$ so as to judge if corrections to scaling have become negligible
rather than going to temperatures down to and below $T_c$. The only
drawback of this approach is that one needs to make an extrapolation in
temperature (as well as in size) in order to estimate $T_c$, the
critical exponents, and the large size universal parameters
$\xi(L,T_c)/L$ and $g(L,T_c)$. 

Cubic lattices of linear size $L=3$ to $L=48$ with bimodal 
random interactions with equal probability and conventional periodic
boundary conditions 
were equilibrated using the exchange Monte-Carlo technique\cite{hukushima:96}. The number of independent samples studied were : 
4000, 4000, 4000, 8192, 8192, 6144, 1024, 2688 and  $512$ for $L=$ 3, 4,
6, 8, 12, 16, 24, 32 and $48$, respectively. Samples up to $L=16$ were
equilibrated 
down to $T=0.9$; in order to avoid the difficulties associated with
excessive equilibration times at the lowest temperatures for large
samples with $L=24, 32$, it was decided to equilibrate only down to
$T=1.1$ and $1.2$ respectively. Two independent batches of $256$ $L=48$
samples were equilibrated down to $T=1.3$ and to $T=1.4$. As discussed above,
for these large sizes this approach avoids the region below the ordering
temperature where equilibration is particularly hard to achieve. 
The observables recorded were the spin-glass susceptibility $\chi(L,T)$, the finite size correlation length $\xi(L,T)$, the Binder parameter $g(L,T)$, and the energy per spin $e(L,T)$. Equilibration was checked by studying the dependence of the observables on the measurement time. Wherever direct comparisons could be made, i.e. for sizes up to $L=24$, there was excellent point by point agreement between the present long time anneal values and equilibrated data from an independent study (Ref.~\onlinecite{katzgraber:06} and Helmut Katzgraber, private communication), confirming that equilibrium had been reached.

\begin{figure}
\includegraphics[width=4in]{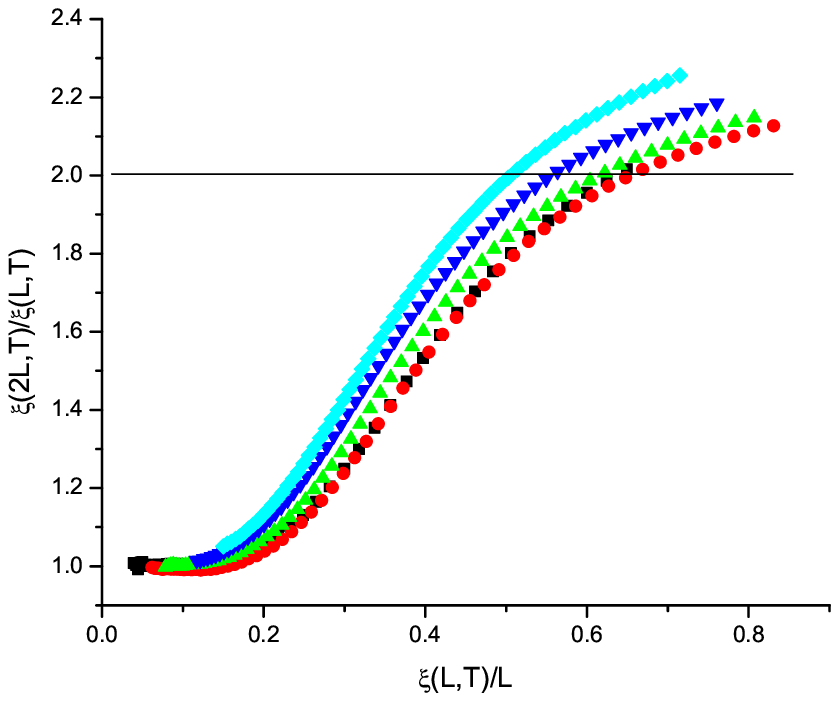}
\caption{(Color online) Scaling plot for correlation length ratios
 $\xi(2L,T)/\xi(L,T)$ against $\xi(L,T)/L$ following
 Ref. \onlinecite{caracciolo:95}. Top to bottom $L=3,4,6,8,12$ cyan,
 blue, green, red, black.} 
\protect\label{fig:1}
\end{figure}

\begin{figure}
\includegraphics[width=4in]{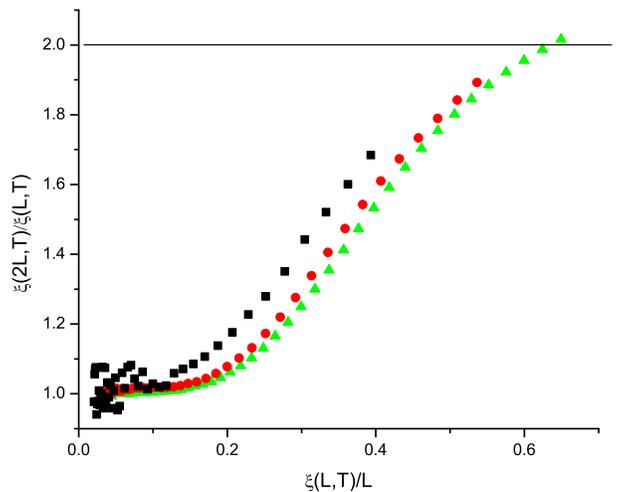}
\caption{(Color online) Scaling plot for correlation length ratios
 $\xi(2L,T)/\xi(L,T)$ against $\xi(L,T)/L$ following
 Ref. \onlinecite{caracciolo:95}. 
Top to bottom $L=24,16,12$ black, red, green.}
\protect\label{fig:2}
\end{figure}

\begin{figure}
\includegraphics[width=4in]{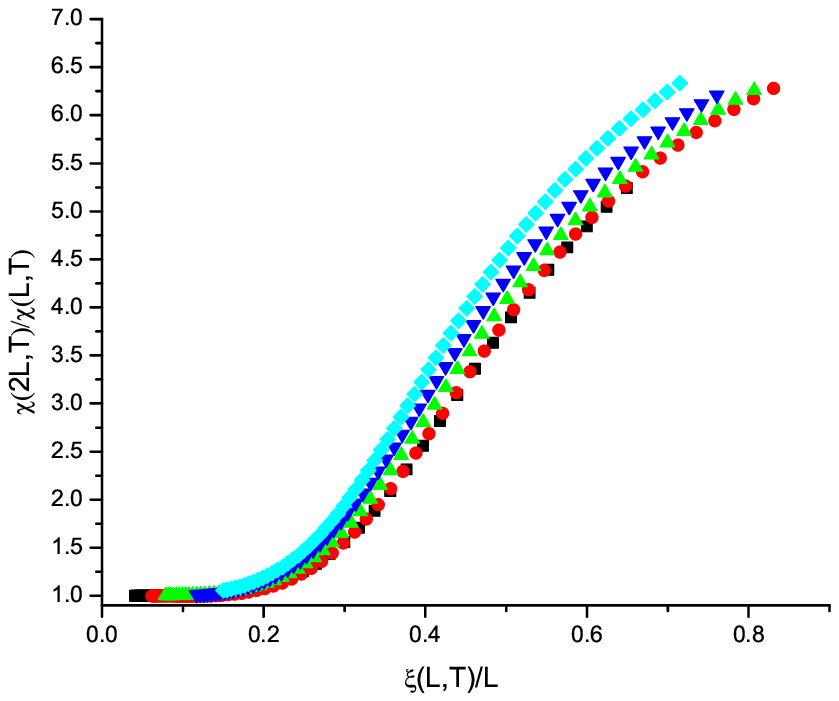}
\caption{(Color online) Scaling plot for reduced susceptibility ratios
 $\chi(2L,T)/\chi(L,T)$ against $\xi(L,T)/L$ following
 Ref. \onlinecite{caracciolo:95}. 
Top to bottom $L=3,4,6,8,12$ : cyan, blue, green, red, black.}
\protect\label{fig:3}
\end{figure}

\begin{figure}
\includegraphics[width=4in]{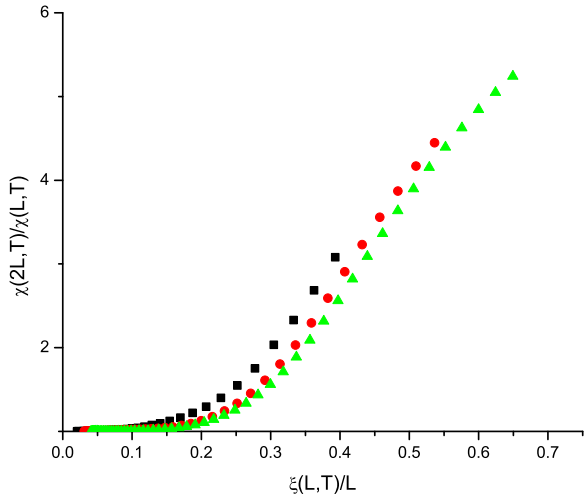}
\caption{(Color online) Scaling plot for reduced susceptibility ratios
 $\chi(2L,T)/\chi(L,T)$ against $\xi(L,T)/L$ following
 Ref. \onlinecite{caracciolo:95}. 
Top to bottom $L=24,16,12$ : black, red, green.}
\protect\label{fig:4}
\end{figure}

\begin{figure}
\includegraphics[width=4in]{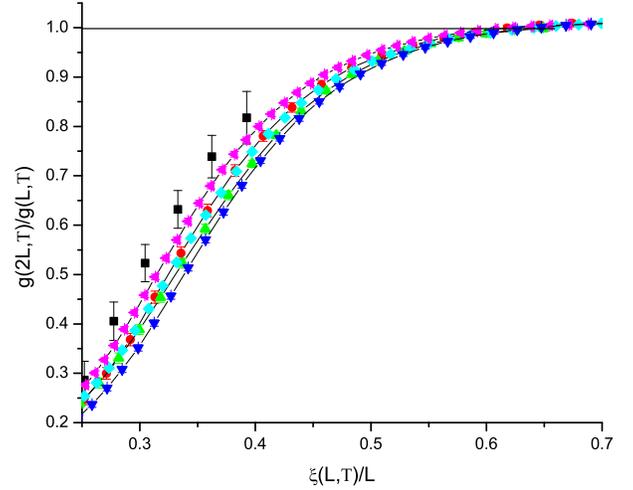}
\caption{(Color online) Scaling plot for Binder parameter ratios
 $g(2L,T)/g(L,T)$ against $\xi(L,T)/L$ following
 Ref. \onlinecite{caracciolo:95}. $L=24, 16, 12, 8, 6, 4$ : black, red,
 green, blue, cyan, magenta.} 
\protect\label{fig:5}
\end{figure}

\begin{figure}
\includegraphics[width=4in]{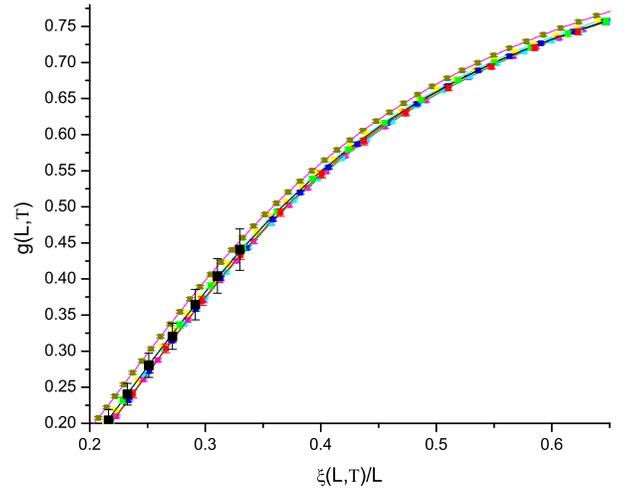}
\caption{(Color online) Scaling plot of the Binder parameter $g(L,T)$
 against $\xi(L,T)/L$. 
$L=48, 32, 24, 16, 12, 8, 6, 4$ : black, red, green, blue, cyan, magenta, yellow, dark yellow.}
\protect\label{fig:6}
\end{figure}

The data were analyzed using the Caracciolo {\it et
al}~\cite{caracciolo:95} finite size ratio scaling
approach. Figures \ref{fig:1} and \ref{fig:2} show the ratio scaling
plots for the correlation length 
$\xi(L,T)$.  In order to be able to visualize the non-monotonic behavior
we present the data in two scaling plots. 
It can be seen that for small $L$ the scaling curves move systematically from left to right with increasing $L$; then as $L$ is increased further beyond $L \sim 24$ the curves begin to move to the left again. 
By inspection the present data
demonstrate that in the 3D bimodal ISG there are significant finite-size
corrections to scaling up to and probably beyond the largest sizes which
we have studied; the observed non-monotonic behavior of the scaling
plots as functions of $L$ implies that there are at least two important
and conflicting correction terms. 
The critical normalized correlation length $(\xi(L,T)/L)_c$ corresponds
to the $x$ coordinate of the intersection of the limiting large size
curve with the horizontal line $y=2$. Data taken only up to $L \sim 24$
give the impression that the large size limit has been reached by this
size, and that $(\xi(L,T)/L)_c \sim 0.65$ (see Katzgraber {\it et al}
and Hasenbusch {\it et al} ~\cite{katzgraber:06,hasenbusch:08}). However
when data up to larger $L$ including $L=48$ are taken into account, it
is clear that one needs to go far beyond $L \sim 24$ to reach the limit where FSS corrections have become negligible; the data indicate that the true large size limit for $(\xi(L,T)/L)_c$ will lie below $0.65$, and will correspond to an ordering temperature $T_c$ in the infinite size limit significantly higher than recent estimates~\cite{kawashima:96,ballesteros:00,katzgraber:06,hasenbusch:08}.

Figures \ref{fig:3} and \ref{fig:4} show the analogous ratio scaling
plot for the SG susceptibility $\chi(L,T)$. Again for small $L$ the curves move
systematically to the right with increasing $L$, and then begin to move
to the left for larger $L$. Once more the position of the infinite $L$
limit curve is difficult to judge but it would correspond to a value for
the exponent $\eta$ less negative than that suggested by recent
estimates~\cite{katzgraber:06,hasenbusch:08}. Figure~\ref{fig:5} shows a ratio
scaling plot for the Binder parameter $g(L,T)$. Again similar behavior
can be observed. It is of interest to note that when the same data are
represented in the form of a plot of $g(L,T)$ against $\xi(L,T)/L$ as in
Fig.~\ref{fig:6}, for $L > 4$ all the points lie on a single curve within the
present statistics. This form of scaling plot is very insensitive to
corrections to scaling because $g(L,T)$ and $\xi(L,T)/L$ are closely
correlated. 

Comparisons have been made between different spin glasses using this  
form of scaling plot (e.g. Ref.~\onlinecite{katzgraber:06}) because systems  
lying in different universality classes should not have the same  
scaling curves. However simple inspection of figures \ref{fig:1}, \ref{fig:2}, \ref{fig:3} and \ref{fig:4}
shows that $g(L,T)$ remains an almost unique function of $\xi(L,T)/L$  
even when strong finite size corrections to scaling are obvious in the  
ratio scaling plots for each of the two parameters taken individually.  
In other words the corrections to scaling for $g(L,T)$ and for  
$\xi(L,T)/L$ are strongly correlated. Because this particular form of  
scaling is very insensitive to corrections, using such plots to judge  
if two systems lie in the same universality class or not would require  
data of extremely high statistical precision.

\begin{figure}
\includegraphics[width=4in]{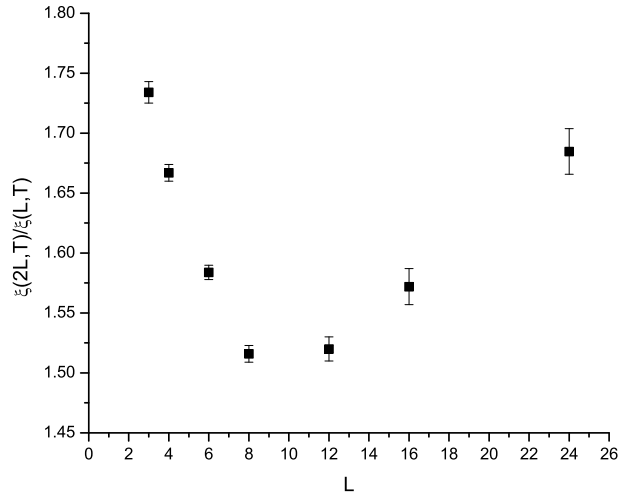}
\caption{Correlation length ratios $\xi(2L,T)/\xi(L,T)$ against $L$ for $\xi(L,T)/L = 0.393$.}
\protect\label{fig:7}
\end{figure}

\begin{figure}
\includegraphics[width=4in]{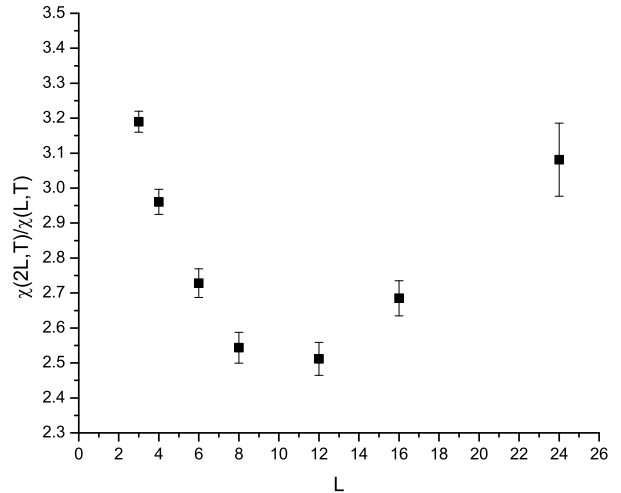}
\caption{Susceptibility ratios $\chi(2L,T)/\chi(L,T)$ against $L$ for $\xi(L,T)/L = 0.393$.}
\protect\label{fig:8}
\end{figure}

\begin{figure}
\includegraphics[width=4in]{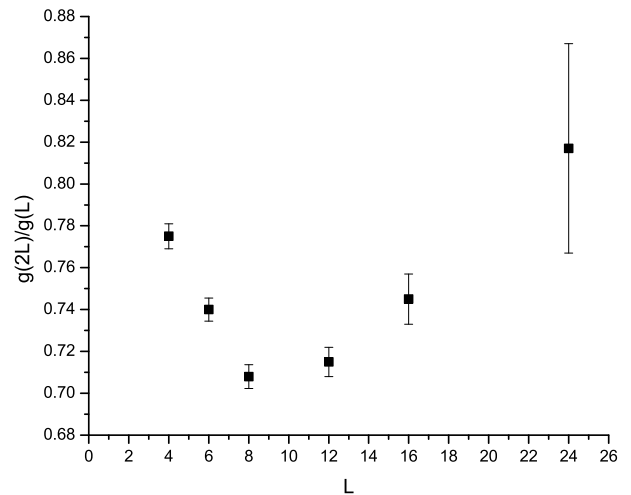}
\caption{Binder parameter ratios $g(2L,T)/g(L,T)$ against $L$ for $\xi(L,T)/L = 0.393$.}
\protect\label{fig:9}
\end{figure}

As an illustration, ratio data are shown as functions of $L$ in figures
\ref{fig:7}, \ref{fig:8} and \ref{fig:9} for one fixed normalized
correlation length value, $\xi(L,T)/L=0.393$. The figures demonstrate
the non-monotonic behavior, and also the difficulty in extrapolating to
infinite $L$ when aiming to estimate the limiting ratio values, even at
temperatures well above criticality. If nevertheless we assume that the
$\xi(48,T)/\xi(24,T)$ ratio curve is close to the infinite size limit,
and we extrapolate the curve to $\xi(48,T)/\xi(24,T)=2$ we obtain $T_c
\sim 1.175$, which in view of the extrapolations we have been obliged to
make can be considered as a lower limit to the true value of
$T_c$. 
We recall that a quite independent ``dynamic'' estimate~\cite{mari:01}
gives $T_c \sim 1.19$. This value is consistent with the 
present equilibrium analysis.

It is important to envisage potential sources of systematic error in the
present data. Even though stringent precautions have been taken to
assure full equilibration, as always with spin glass simulations a
possible source of error which must be kept in mind would be incomplete
equilibration. As stated above, the data for sizes up to $L=24$ are in
very good agreement with those from reliable independent studies(Ref.~\onlinecite{katzgraber:06} and Helmut Katzgraber, private communication). It
should be noted that a lack of equilibration for the larger samples only
($L=48$ and $L=32$) would reduce $\xi(2L,T)$ and $\chi(2L,T)$ for $L=24$
and $L=16$ while leaving $\xi(L,T),\chi(L,T)$ and $\xi(L,T)/L$
unaltered, and so would artificially move the scaling curves of Figures
2 and 4 downwards. The observation that the curves for
$\xi(48,T)/\xi(24,T)$ and $\xi(32,T)/\xi(16,T)$ in these two figures lie
well above the others thus cannot be an artifact arising from lack of
equilibration at $L=48$ and $L=32$. There remain of course purely
statistical sampling effects due to the fact that at large sizes the
number of samples studied is necessarily restricted. For the sizes
$L=48$ and $L=32$ the only data with which comparisons could be made are
those from the pioneering work of Ref.~\onlinecite{palassini:99}; these are
fragmentary and recorded on far fewer samples than in the present work,
but within the statistical errors there is agreement, particularly for
the two $L=48$ points from $64$ samples recorded by
Ref.~\onlinecite{palassini:99}.  

For the 3D Gaussian ISG, corrections to scaling are much very weaker
than in the bimodal case~\cite{katzgraber:06}, so it can plausibly be
assumed that the estimates for $T_c$ and for the critical values of
$\xi(L,T)/L$ and $g(L,T)$ from the largest $L$ Gaussian ISG
data~\cite{katzgraber:06} can be taken as being essentially equal to the
infinite size thermodynamic limit values. Unless strong corrections
appear when still larger $L$ Gaussian samples are studied, the Gaussian
critical parameters in the thermodynamic limit are not equal to those
for the thermodynamic limit bimodal ISG derived as outlined above,
implying that the two systems do not lie in the same universality
class.

\section{Finite-size corrections to scaling}

Clearly it is essential to allow for corrections to scaling in order to obtain reliable estimates of the critical parameters.
At criticality FSS expressions including correction terms can be written
as (see for instance Ref.~\onlinecite{hasenbusch:06})
\begin{equation}
\chi(L,\beta_c) = C_{\chi}L^{2-\eta}[1 + a_{1}L^{-\omega} + a_{2}L^{-\omega_{2}} +\ldots]
\end{equation}
where $\omega$ and $\omega_{2}$ are the first and second correction exponents.
For standard ferromagnets these exponents are rather accurately known from RGT; in dimension 3 $\omega \sim 0.51$ and $\omega_{2} \sim 0.82$ in pure systems and  $\omega \sim 0.3$ and $\omega_{2} \sim 0.8$ ~\cite{hasenbusch:06} for diluted ferromagnets. Unfortunately there are no equivalent RGT guide-lines for the values of the correction exponents (and {\it a fortiori} for the strengths of the correction terms) in ISGs.
The non-monotonic variation of the positions of the scaling curves with
increasing $L$ in the 3D bimodal ISG implies that there are two
significant contributions (at least) with opposite signs to the
finite-size-scaling corrections, coming from $\omega$ and $\omega_{2}$
terms. 

For finite-size scaling with corrections near $T_c$, 
Calbrese {\it et al} ~\cite{calbrese:03} give the ansatz
\begin{equation}
\xi(2L,T)/\xi(L,T) = F(L/\xi(L,T)) + L^{-\omega}G(L/\xi(L,T))
\label{calabrese}
\end{equation}
and the equivalent equation for $\chi$, where $F$ and $G$ are scaling functions and $\omega$ is the leading non-analytic correction to scaling exponent. Unfortunately the present data are manifestly insufficient to attempt fits to analogous equations with two correction terms, so we can make no sensible estimate of the correction exponents in the 3D bimodal ISG.

It can be noted that in the 2d Bimodal ISG (which orders only at $T=0$) limiting behavior at criticality is only attained above $L \sim 100$~\cite{hartmann:01}.

\section{Conclusion}

Equilibrium numerical data on the three dimensional bimodal Ising spin
glass for samples up to size $L=48$ show that corrections to finite-size
scaling are unexpectedly strong and non-monotonic. Taking into account
the influence of these corrections, it can be concluded that the large
size limit behavior and the corresponding thermodynamic limit critical
temperature and critical parameters are significantly different from the
values derived from previous equilibrium simulation data where smaller
maximum sizes were studied. Estimates of critical parameters from the
present data are only indicative due to the intrinsic problem of
extrapolating  reliably to infinite size, but they are consistent with
those obtained from a ``dynamic''
method~\cite{bernardi:96,mari:01,pleimling:05}. 
The strong corrections to finite size scaling must be fully taken into  
account before a valid comparison can be made between the  
thermodynamic limit behaviors for the bimodal and Gaussian 3D ISGs.  
The data as they exist cannot be considered to show that the two systems  
lie in the same universality class.

\begin{acknowledgments}
We would like to thank Helmut Katzgraber for generously making his
numerical data available to us, and for numerous helpful comments. We
would also like to thank Andrea Pelissetto for helpful explanations of
RGT FSS theory.
This work was partly supported by the Grant-in-Aid for Scientific
Research (No. 1807004). 
The numerical calculations were mainly performed on the 
facilities of the Supercomputer Center, Institute for Solid
State Physics, University of Tokyo.
\end{acknowledgments}

\end{document}